\documentclass[aps,prb,preprint,onecolumn,citeautoscript]{revtex4-1} 
\usepackage{float}
\usepackage{amsmath,amssymb,mathrsfs,bm,feynmf,setspace}
\usepackage{graphicx}
\usepackage{color}
\usepackage{mathtools}
\usepackage{verbatim} 
\usepackage[papersize={8.5in,11in}]{geometry}

\usepackage[colorlinks=true]{hyperref} 
\hypersetup{
    bookmarks=true,         
    unicode=false,          
    pdftoolbar=true,        
    pdfmenubar=true,        
    pdffitwindow=false,     
    pdfstartview={FitH},    
    pdftitle={My title},    
    pdfauthor={Author},     
    pdfsubject={Subject},   
    pdfcreator={Creator},   
    pdfproducer={Producer}, 
    pdfkeywords={keyword1} {key2} {key3}, 
    pdfnewwindow=true,      
    colorlinks=true,       
    linkcolor=red,          
    citecolor=blue,        
    filecolor=magenta,      
    urlcolor=cyan           
}
\newcommand{\be}{\begin{equation}}
\newcommand{\ee}{\end{equation}}
\newcommand{\bea}{\begin{eqnarray}}
\newcommand{\eea}{\end{eqnarray}}
\newcommand{\tr}{\textrm{tr}}
\def\bse{\begin{subequations}}
\def\ese{\end{subequations}}

\begin{document}

\title{Revisiting Entanglement Entropy of Lattice Gauge Theories}
 \author{Ling-Yan Hung}
 \email{lyhung@fudan.edu.cn}
\affiliation{Department of Physics and Center for Field Theory and Particle Physics, Fudan University, Shanghai 200433, China}
\affiliation{Collaborative Innovation Center of Advanced Microstructures, Nanjing University, Nanjing , 210093, China.}
 \author{Yidun Wan}
 \email{ywan@perimeterinstitute.ca}
 \affiliation{Perimeter Institute for Theoretical Physics, Waterloo, ON N2L 2Y5, Canada}

\begin{abstract}
Casini et al raise the issue that the entanglement entropy in gauge theories is ambiguous because its definition depends on the choice of the boundary between two regions.; even a small change in the boundary could annihilate the otherwise finite topological entanglement entropy between two regions.  In this article, we first show that the topological entanglement entropy in the Kitaev model \cite{Kitaev2003a} which is not a true gauge theory, is free of ambiguity. Then, we give a physical interpretation, from the perspectives of what can be measured in an experiement, to the purported ambiguity of true gauge theories, where the topological entanglement arises as redundancy in counting the degrees of freedom along the boundary separating two regions. We generalize these discussions to non-Abelian gauge theories.   
\end{abstract}
\maketitle

\section{Introduction}
Casini \textit{et al} explores the issue of defining entanglement entropy for gauge theories. Due to gauge redundancy, it is subtle to decide how to split the Hilbert space into products of subspaces each containing states within a given spatial region\cite{Casini2013}. To make the question precise, Casini chooses to define a Hilbert space by the operator algebra that acts on the space. Having chosen the algebra $\mathcal{A}_V$ defining the Hilbert subspace of region $V$, its commutant $\mathcal{A}_V' \equiv \mathcal{A}_{\overline{V}}$ is thus the operator algebra that defines the Hilbert subspace of the complement region $\overline{V}$. It is generally possible that $\mathcal{A}_V$ and $\mathcal{A}_{\overline{V}}$ share a common center $\mathcal{C}$, in which case the Hilbert space cannot be split into a product of two subspaces acted on by $\mathcal{A}_{V}$ and $A_{\overline{V}}$ separately. When such a common center is non-trivial, the computation of the entanglement entropy between $V$ and $\overline{V}$ has to be properly defined. It is argued that the most natural way is to express the density matrix $\rho$ in the basis such that the common center $\mathcal{C}$ is diagonalized. One then throw away all off-diagonal components in $\rho$ that maps states with different eigenvalues with respect to $\mathcal{C}$, and each of the diagonal blocks characterized by given eigenvalue of $\mathcal{C}$ would now have a clear product structure of states acted on by $\mathcal{A}_{V}$ and $A_{\overline{V}}$, from which we can obtain blocks of reduced density matrix by tracing out states in $\mathcal{A}_{\overline{V}}$. The final entanglement entropy would thus take the form as a sum of two parts, the classical part and the quantum part, namely,  
\be
S_{E} = - \sum_i p_i \ln p_i - \tr (\rho^i_V\ln \rho^i_V),
\ee
where $p_i$ is the overall coefficient of a given diagonal block in the density matrix $\rho$.

It was questioned how the choice of $\mathcal{A}_{V}$ affects the value of $S_{E}$, particularly in the case where the ground state concerned is a topological phase characterized by long-range entanglement, and with a distinct topological entanglement entropy in the case of 2+1 dimensional topological ordered phases\cite{Kitaev2006a,Levin2004}. This question is explored specifically in the context of 2+1 dimensional lattice gauge theories. In the specific examples of Kitaev models corresponding to Abelian groups, it is shown that one can readily find choices of $\mathcal{A}_{V}$ for which the entanglement entropy defined above is identically zero, and that subsequently the topological entanglement entropy is also zero. We would like to elucidate the subtleties in this problem. 

\section{Entanglement entropy in the Kitaev model}
Let us verify that the Kitaev models have the usual
topological entanglement entropy, quite independent of the choice of
$\mathcal{A}_{V}$.
\subsection{Electric vs magnetic center in the Kitaev model}
The Kitaev model defined in Ref.\onlinecite{Kitaev2003a} is not truly a gauge theory, in the sense that there is not any gauge redundancy in the Hilbert space, except that the ground state is chosen to be within the gauge invariant subspace, satisfying
\be
A^g_{v_i}|\psi\rangle = |\psi\rangle,
\ee
where $A^g_{v_i}$ are the star operator defined as in Ref.\onlinecite{Kitaev2003a}.
\be \label{gaugetrans}
A^g_{v_i}= \prod_{l \in v_i} L^g_l,
\ee
where $L^g_l$ acts on the link variable on the link $l$ and multiply the link
variable by the group element $g$, and $\l\in v_i$ refers to all the links meeting at the vertex $v_i$. Whether it is a left or right multiplication depends
on whether an oriented link $l$ points toward or away from the vertex $v_i$. The convention chosen in Ref.\onlinecite{Kitaev2003a} is such that

\be
L^g_l |h\rangle_{\textrm{in}} = |gh \rangle_{\textrm{in}}, \qquad L^g_l |h\rangle_{\textrm{out}} = |hg^{-1} \rangle_{\textrm{out}}
\ee
and the state on a link $|h\rangle_{\textrm{in,out}}$ is defined where $h \in G$.

The idea of the electric center and the magnetic centers are basically two natural
choices of $\mathcal{A}_{V}$ in a lattice gauge theory with gauge group $G$.
Consider a set of vertices $V\equiv \{v_i\}$.
The electric centre corresponds to choosing a
\emph{rough boundary} for region $V$, namely that one includes all the operators $L^g_l$ generating
group multiplication on the link variables $l$ 
that connects to at least one vertex $v$ in region $V$.
\begin{figure}[ht]
\centering
\includegraphics[scale=2]{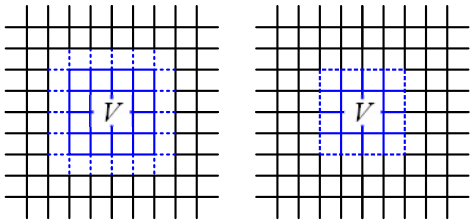}
\caption{(Color online) Electric (rough) boundary on the left and magnetic (smooth) boundary on the right. The boundaries are in dashed lines, which are part of region $V$.}
\label{fig:EMboundaries}
\end{figure}

\subsection{Naive entanglement entropy with the magnetic center }

Before we move on to computing entanglement entropy based on the
choice of the operator algebra, let us consider computing the entanglement entropy in
the most naive way, where we separate the links into 2 sets, without cutting links as being usually
considered to preserve the original lattice structure. Consider in particular
a convex region surrounded by a rough boundary. 

We recall that the ground state of the Kitaev model is given by
\be\label{eq:GSdef}
|\psi\rangle = \mathcal{N} \prod_{v_i} (\frac{1}{|G|}\sum_g  A^{v_i}_{g}) |0\rangle, 
\ee 
where $|0\rangle$ is the `empty' state in which the degree of freedom on each link is set to the identity element of the gauge group $G$, and the factor  $\mathcal{N}$ in the front is to ensure correct normalization. We have
\be
\mathcal{N} = |G|^{v_i/2}.
\ee

We can then rewrite the above ground state as follows
\be
|\psi\rangle =  \sum_{i,j,k} X^{\{a_i,b_i,c_i\}} |\{c_i\}\rangle \otimes |\{b_j\} \rangle \otimes |\{a_k\}\rangle,
\ee
where $\{c_i\}$ denotes the set of configurations of the links in region $\bar{A}$, the complement of region $A$,
$\{a_i\}$ is the set of configurations inside region A except for links along the rough boundary, and
$\{b_i\}$ configurations inside the rough boundary. Note that $i,j,k$ can be composite indices, as to be seen soon. Coefficients $X^{\{a_i,b_i,c_i\}}=1$
if a configuration is allowed, and zero otherwise.
\begin{figure}[ht]
\centering
\includegraphics[scale=3]{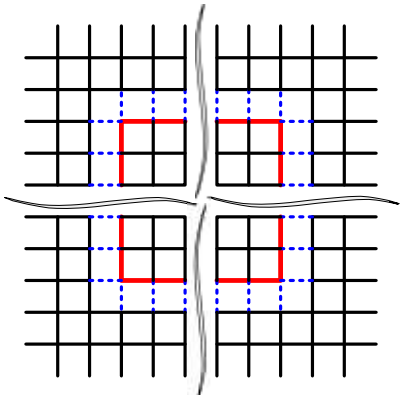}
\caption{(Color online) An infinite lattice divided into two regions: $A$, inside the red (thicker line) loop, and $\bar A$, outside of the loop. the blue (dashed) lines comprising the rough boundary also belong to $A$. The wiggly lines simply represent the parts of the infinite lattice that are neglected.}
\label{fig:roughBC}
\end{figure}

Consider the lattice in Fig. \ref{fig:roughBC}, which is divided into two regions $A$, inside the red (thicker) loop and including the rough boundary  (blue dashed) links, and $\bar A$. Suppose there are exactly $L$ links comprising the rough boundary, and the numbers of sites outside and inside the red (thicker) loop are respectively $m$ and $n$, and let $M=|G|^{m-L}$ and $N=|G|^{L-4-1}$. We can rearrange the ground state as follows (and omit the curly brackets when necessary to avoid clutter):
\bea\label{eq:GSnaive}
|\psi\rangle =& & |\{c_1\}\rangle \otimes \left[\sum_g \left( |  b^g_{11}\rangle\otimes |\{ a_1\}\rangle + 
|b^g_{12}\rangle \otimes |\{a_2\}\rangle + \cdots |b^g_{1N} \rangle\otimes |\{a_{N}\}\rangle  \right) \right]\nonumber \\
&&+ |\{c_2\}\rangle \otimes \left[\sum_g \left( |  b^g_{21}\rangle\otimes |\{ a_1\}\rangle + 
|b^g_{22}\rangle \otimes |\{a_2\}\rangle + \cdots |b^g_{2N} \rangle\otimes |\{a_{N}\}\rangle  \right) \right]\nonumber \\
&&+ |\{c_3\}\rangle \otimes \left[\sum_g \left( |  b^g_{31}\rangle\otimes |\{ a_1\}\rangle + 
|b^g_{32}\rangle \otimes |\{a_2\}\rangle + \cdots |b^g_{3N} \rangle\otimes |\{a_{N}\}\rangle  \right) \right]+ \cdots\nonumber \\
&& + |\{c_{|G|^{L-1}}\}\rangle \otimes \left[\sum_g \left( | b^g_{|G|^{L-1}1}\rangle\otimes |\{ a_1\}\rangle 
+ \cdots |b^g_{|G|^{L-1}N} \rangle\otimes |\{a_{N}\}\rangle  \right) \right]
\eea
where each $|\{c_i\}\rangle$ does not correspond to a single configuration but rather the sum of $|G|^{m-L}$ different configurations in $\bar A$, and each $|\{a_i\}\rangle$ corresponds to the sum of $|G|^{n-L+4}$ different configurations inside region $A$. It is clear that there are precisely $|G|^{m+n-1}$ terms in the above equation, in accordance with Eq. \eqref{eq:GSdef}. 

The reason it takes the above form is that the ground state contains all linear combination of possible
ways of acting $A^{v_i}_g$ on $|0\rangle$, leading to $|G|^{m+n-1}$ different configurations of $G$ elements on the lattice links. We can group all possible configurations of the links in $\bar A$ by the configurations on the links on the rough boundary. The $L$ links on the rough boundary give rise to exactly $|G|^{L-1}$ boundary configurations, each of which is associated with precisely $|G|^{m-L}$ configurations in $\bar A$ because there are $m-L$ sites in $\bar A$. Therefore, each $|\{c_i\}\rangle$ is the sum of $|G|^{m-L}$ configurations in $\bar A$ that share the same configuration of the rough boundary. This way, the configurations inside the red loop can be considered separately. Now for each $|\{c_i\}\rangle$, the group elements on all the links on the rough boundary are fixed at once; hence, regardless of the detailed group values on these links, there would be exactly $|G|$ possible simultaneous choices of these values, resulting in the $\sum_g|b_{ij}\rangle$
in Eq. \eqref{eq:GSnaive}. Consequently, for each rough boundary configuration, namely for each $|b_{ij}\rangle$, we can similarly group the configurations in $A$ excluding the rough boundary by  the  $A^{v}_g$ actions on the sites $v\in A$ that are directly at the end of the links on the rough boundary, and there are just $L-4$ such sites. Thus, we have $N$ groups of $|\{a_{1\leq j\leq N}\}\rangle$ configurations, each of which is the sum of $|G|^{n-L+4}$ configurations in $A$ excluding the sites right at the rough boundary, the number of which is exactly $n-L+4$.

If we now obtain the reduced density matrix, in which we trace out $|c_i\rangle$, the vectors inside the big bracket are clearly orthogonal eigenstates of the reduced density matrix. 
Since there are exactly $|G|^{L-1}$ such eigenstates, and that each orthogonal eigenvector can be normalized and the eigenvalues sum to 1, the von-Neumann entropy is 
\be
S_{E} = \ln |G|^{L-1} = L \ln |G| - \ln |G|.
\ee
We thus recover the topological entanglement entropy $\gamma = \ln |G|$ which is the quantum dimension of the Kitaev model. 

In the Kitaev model, had we chosen to have smooth boundary instead of rough boundaries, the same arguments can be run through using the projector operator $B_g$. Since there is exact electric-magnetic duality, the result is independent of such choices of boundaries.

When there are concave boundaries, then the number of vertices outside of  $A$ that controls
the rough boundary would be less than the number $L$ of boundary links on the rough boundary.
However, that number depends on the precise form of the boundary and the number of vertices around the rough boundary outside of $A$ scales linearly with $L$. Therefore, in the thermodynamic limit, we have a stable answer for the topological entanglement entropy that is truly $L$ independent.

\section{Entanglement entropy in ``true'' gauge theories vs SPT}
As noted in the previous subsection, we see that the topological entanglement entropy does not depend on the electric/magnetic boundaries for the Kitaev model. We would like to understand
here why the calculation in Ref.\onlinecite{Casini2013} has led to vanishing topological entanglement entropy for magnetic centers and in fact the same is true for a \emph{maximal tree} structure in a true gauge theory. 
A key characteristic in the representation of the Hilbert space of a gauge theory is the redundancy in the labeling of the states: two configurations related by a gauge transformation generated by (\ref{gaugetrans}) are one and the same. Instead of claiming that the ground states are invariant under (\ref{gaugetrans}) by summing over all the states generated by (\ref{gaugetrans}) acted on different number of vertices, one requires that each orbit of the gauge transformations corresponds to one single state. 

Let us take the most extreme example in which we consider a single spin 1/2 system, in a 2d Hilbert space.
\be
H= \{|\uparrow\rangle, \,\, |\downarrow\rangle \}.
\ee
Now suppose we just add a redundant label to the above states
\be
H= \{|\uparrow, \uparrow\rangle, \,\, |\downarrow, \downarrow\rangle \},
\ee
which clearly is completely trivial. Consider the state
\be
|\Psi\rangle = \frac{1}{\sqrt{2}} |\uparrow, \uparrow\rangle +  \frac{1}{\sqrt{2}} |\downarrow, \downarrow\rangle.
\ee
Let us take $\sigma^1_{x,y,z}$ as Pauli matrices acting on the ``first'' spin.
While there is not really a  second spin degree of freedom, we can always make the claim that
$\sigma^2_z$ that acts on the ``second spin'' takes eigenvalues satisfying
\be \label{gauge_condtri}
\sigma^1_z \times \sigma^2_z = 1,
\ee
which is the direct analogue of gauge conditions in a gauge theory, although in a much more transparent form here.
Now using basis diagonalizing $\sigma^1_z$, the density matrix is
\be
\rho = |\Psi\rangle\langle\Psi| = \frac{1}{2} \left(
\begin{array}{cc}
1 & 1\\
1&1
\end{array}
\right)
\ee
which is a two by two matrix, unlike the density matrix of a true 2 spin systems, which would have been a 4 by 4 matrix. In this case, is there a sense in which there is any entanglement? Clearly, there is not. However, entanglement entropy could also be viewed as the ignorance accompanying the dumping of some operators from the set of observables which we measure. i.e. the inability to determine the precise form of the density matrix if we only have measurements of a subset of all the observables. So one could talk about the density matrix, mostly likely mixed, with maximal entropy that reproduces the measurements of the remaining observable.
So for instance, if we are not allowed to make measurements of $\sigma^1_x$ or $\sigma^1_y$, then clearly the above density matrix cannot be distinguished from the maximally mixed state
\be
\rho_{\textrm{max}} =   \frac{1}{2} \left(
\begin{array}{cc}
1 & 0\\
0&1
\end{array}
\right),
\ee
which has entropy $\ln 2$. This is precisely the same entropy we would attribute to the two spin system had we taken the second spin to be a true extra degree of freedom, although it is clearly not the standard meaning of entanglement entropy. But had we started with two real spin degrees of freedom, we can also think of the entropy resulting from tracing out the second spin as the ignorance introduced when we stop making measurements of operators of the form $\mathcal{O}^1 \otimes \sigma^2_{x,y,z}$, where $\mathcal{O}^1 \in \{1_2,\sigma^1_{x,y,z}\}$. That is, we are only allowed to make measurement of the first spin using operators $\mathcal{O} \in \{\sigma^1_{x,y,z}\otimes 1_2\}$. This is of course the familiar notion
of keeping only local knowledge within a chosen ``region''. 
 What we attempt to illustrate with this trivial example are two points:
first, that entanglement can be viewed as ignorance after dumping some observables from being measured; second, which trivially follows from the first assertion, is that the choice of measurable observables directly affects the final value of the entropy. This is precisely the point made in Ref.\onlinecite{Casini2013} over the 
effect of the choice of the operator algebra $\mathcal{A}_V$ on the value of the entanglement. 
The entanglement entropy following from the electric center choice is the ignorance introduced from dumping the loop magnetic operators at the boundary of the region. This is the direct analogue of the above entropy for the single spin system where we dump $\sigma_x^1$ from our measured set. To reiterate, the procedure of \emph{deleting} off-diagonal components in the density matrix after expressing it in the eigenbasis of the electric link operators $L^g_l$ is indeed simply equivalent to dumping magnetic operators non-local with the boundary $L^g_l$. Whether this is considered \emph{classical} contribution to the entropy, as asserted by Casini is perhaps more or less a matter of taste.

As has been emphasized in Ref.\onlinecite{Casini2013} therefore, an ambiguity, or more precisely, some artificial choice one has to make in defining the entanglement entropy is not only special to gauge theories, but to all kinds of theories, when we make different choices of sets of operators that we perform measurements. The major difference of a gauge theory compared to a local scalar field theory however, is that there exists, in the case of say the local scalar field theory, a natural or canonical choice of the operators that we keep, which defines a notion of spatial entanglement, and that such a natural choice is absent in a gauge theory.  

The crucial question, therefore, boils down not so much to which is a \emph{correct} choice of operator algebra $\mathcal{A}_V$, but rather, which is a \emph{useful} choice that would give us useful information about the system under scrutiny. 

Here, we recall that entanglement entropy (or more precisely, mutual information) is a measure of correlations between different degrees of freedom. Of course, strictly speaking the correlation following from (\ref{gauge_condtri}) is trivial -- states not satisfying the condition simply does not exist. Nevertheless, realistically, in an experimental setting, it is not the Hilbert space that we are directly probing, but the expectation values of (gauge invariant) observables. Therefore, from the perpsective of an experiementalist, which do not have a clue of the redundancy of the Hilbert space, would be fascinated by the agreement between the results of $\sigma^1_z$ and $\sigma^2_z$. Therefore it is perhaps too immature to discard the idea that the entanglement entropy of the electric center is just as real an entanglement entropy as it could be. 

We note also that the dramatic difference between the electric centre and magnetic centre simply follows from the fact that the ground state behaves as a direct product state if we use a basis that consists of the eigenstates of the magnetic flux loop operator. We note that this is precisely the same basis for the $\mathbb{Z}_2$ symmetry protected topological (SPT) phase, recalling the ``duality'' relation between SPT and gauge theory as first described in Ref.\onlinecite{Levin2012} and elucidated in detail in Ref.\onlinecite{Hung2012}. It is therefore not surprising that the topological entanglement entropy should vanish.

\section{Generalization to Non-Abelian theories}
In non-Abelian gauge theories, the degrees of freedom again live on edges, and are labeled by group elements of a gauge group $G$.  One needs a convenient representation of the gauge invariant Hilbert space in order to discuss the splitting of space. This is well-known in tensor network theory. The idea is that one can perform a discrete generalized Fourier transform on the states on each edge, as follows \cite{Buerschaper2009}
\be
|\mu, a,b\rangle = \sum_g  \sqrt{\frac{\mu}{|G|} } \rho^\mu_{ab}(g) |g\rangle,
\ee
whereas the inverse transform is given by
\be
|g\rangle = \sum_\mu \sqrt{\frac{\mu}{|G|} } \rho^\mu_{ab}(g)^* |\mu, a ,b\rangle.
\ee
Note that we abused the notation $\mu$ for both an irreducible representation of $G$ and the dimension of the representation.

Under a gauge transformation therefore, the new basis transforms as
\bea
L^g_l |h\rangle_{\textrm{in}} = |gh\rangle_{\textrm{in}} \,&& \to L^g_l |\mu, a, b\rangle_{\textrm{in}} = \rho^{\mu}(g^{-1})_{a c} |\mu, c, b\rangle_{\textrm{in}} \nonumber \\
L^g_l |h\rangle_{\textrm{out}} = |hg^{-1}\rangle_{\textrm{out}} \,&& \to L^g_l |\mu, a, b\rangle_{\textrm{out}} =  |\mu, a, c\rangle_{\textrm{out}}\, \rho^{\mu}(g)_{c b},
\eea
where repeated indices are summed.

The ground state can now acquire a gauge invariant representation as we translate the ground state into this representation. Recall that the ground state of the theory is given by equation \eqref{eq:GSdef}. While we have emphasized that configurations related by gauge transformation in a gauge theory is strictly identified, one could think of \eqref{eq:GSdef} as a non-gauge-fixed representation, to make the ground state appear explicitly gauge invariant.

Now, if we just follow through the change of basis, the ground state would look like the following
\be
|\psi\rangle = \mathcal{N}\prod_{v\in \textrm{vertices}} (\frac{1}{|G|}\sum_g A^v_g) \prod_{l\in \textrm{links}} (\sum_{\mu_l, a_l} \sqrt{\frac{\mu_l}{|G|}} |\mu_l, a_l, a_l\rangle_{\textrm{appropriate orientation}}).
\ee 

It's not hard to understand what the projectors do. To make it explicit, we focus, without loss of generality, on a particular 3-point vertex, and let us assume that all links are pointing outwards. Then the action of the projector gives 
\be
|\psi\rangle =  \frac{1}{\sqrt{|G|}} \sum_g \sum_{\{\mu_i, c_i, \cdots\}}\sqrt{\frac{\mu_1\mu_2\mu_3}{|G|^3}} |\mu_1,a_1,c_1\rangle
 |\mu_2,a_2,c_2\rangle  |\mu_3,a_3,c_3\rangle \rho^{\mu_1}(g)_{c_1b_1}\rho^{\mu_2}(g)_{c_2b_2}\rho^{\mu_3}(g)_{c_3b_3} \cdots ,
\ee
where 
\be
\sum_g  \rho^{\mu_1}(g)_{c_1b_1}\rho^{\mu_2}(g)_{c_2b_2}\rho^{\mu_3}(g)_{c_3b_3} 
= \frac{|G|}{\mu_3} (\omega^{\mu_1\mu_2 \bar{\mu_3}}_{c_1 c_2 c_3})^\dag \omega^{\mu_1\mu_2 \bar{\mu_3}}_{b_1 b_2 b_3} ,
\ee where the $\omega^{\{\mu_i\}}_{\{c_i\}}$ are the intertwiners that map product representations of $\mu_1\otimes \mu_2$ to representation $\bar{\mu_3}$. The factor of $1/\mu_3$ follows from the identity
\be
\frac{1}{|G|} \sum_g \rho^{\mu}(g)_{a b} \rho^{\mu}(g^{-1})_{c d} = \frac{1}{\mu} \delta_{a d}\delta_{b c}.
\ee 
The statement is simply that by summing over $g$, the surviving terms correspond to the singlet representation by decomposing the product representation of $\mu_1 \otimes \mu_2 \otimes  \mu_3$, and for simplicity we assume there is only one fusion channel to the singlet. Otherwise we will have multiple versions of $\omega$, each corresponding to a particular fusion channel. Had the vertex been endowed with a valency greater than 3, the end product would be a series of fusions that take the representation to the singlet representation. 
Therefore, we have
\be \label{gs1}
|\psi\rangle = \sqrt{G}\sum_{\mu_i, a_i,b_i,c_i} \sqrt{\frac{\mu_1\mu_2}{|G|^3 \mu_3}} 
 (\omega^{\mu_1\mu_2 \bar{\mu_3}}_{c_1 c_2 c_3})^\dag \omega^{\mu_1\mu_2 \bar{\mu_3}}_{b_1 b_2 b_3} |\mu_2, a_2, c_2\rangle |\mu_1, a_2, c_2\rangle |\mu_3, a_3, c_3\rangle\cdots ,
\ee
and if the system consists of this single vertex, we shall have $a_i= b_i$ that are also summed over but will not have any term in the neglected part '$\cdots$'. We thus recover the well-known fact in tensor-network theories, that the gauge invariant Hilbert space consists of all configurations specified by a set of $\{\mu_i\}$ at each links, such that when multiple links meet at a vertex, there must exist a fusion channel such that they fuse to the trivial representation. If more than one such fusion channel exists, we would have to specify the particular channel. Therefore a state is specified by  
\be|\Psi\rangle = |\{\mu_i\}, \{\gamma^v_i\}\rangle,\ee
where we take $\gamma^{v_i}$ to denote the particular fusion channel to the singlet 
at a given vertex $v_i$. The internal indices are not gauge invariant; hence, they are not needed to specify the Hilbert space. 

Now the question is, if we were to cut the system into two pieces in some sense, how would we attribute entanglement entropy to these states? On the one hand, the internal indices are not gauge invariant and thus unphysical. On the other hand, we have no other way to specify the states without referring to them, and the particular contraction of the link internal indices across vertices again make it ambiguous to define a product space, and it appears that we are no better off than the previous basis. At this point, we can draw experience from our discussion in the previous section and also the case of Abelian theories. Recall in the case of Abelian theories, the outcome of measurements of ``spins'' connected by a vertex are correlated due to the gauge invariant constraint. While we argued that such a entanglement might be trivial, it is still of interest to quantify these observed correlation, by discarding a set of observables from the set of observables that would be measured, and ask for the density matrix with the largest amount of entropy that would agree with the original density matrix over the set of measured observables. 
Phrased in these terms, it is clear what measurements we would like to keep making. We would like to be able to measure $\mu_i$, which are gauge invariant variables, and we would find uncanny correlation in these measurements of these representation labels $\mu_i$ when the links meet at a vertex. 
As in the case of picking electric center of Abelian theories, any such ``entanglement'' entropy has to arise at the boundary, where we give up measuring observables not commuting with the measurement of $\mu_i$. In non-Abelian theories, this would again boil down to Wilson loops, much like what happens in Abelian theories.  To compute the resultant entropy that characterizes our ignorance, we would like to start with computing the probability density of obtaining a particular set of outcomes $\{\mu\}$ for links taken to be in region $A$. In the interior region, these outcomes have no dependence on dofs in $\bar{A}$, and only at the boundary they are controlled by links in $\bar{A}$ via Gauss's law. Once we obtain the probability density of individual outcomes for these boundary links (picture exactly like the electric center in \cite{Casini2013}), we would then be able to recover the entropy by 
\be
S= - \sum_ip\{\mu_b\}_i \ln p\{\mu_b\}_i,
\ee
and this would be understood as the entanglement entropy. Again let us emphasize this is precisely the same procedure as in Ref.\onlinecite{Casini2013} of obtaining a density matrix prepared in a basis that diagonalizes the electric observables, and then ``deleting'' off diagonal entries before computing a ``classical'' entropy arising from the distinct blocks in the resulting mixed density matrix.

Let us begin with a simpler example, in which the whole system consists of again precisely one vertex connecting three links. We will take one link, labeled by $\mu_3$, as belonging to region $A$, and the other two to $\bar{A}$. 
The entanglement of integrating out $\bar{A}$ can be computed using the form of the ground state as given in (\ref{gs1}).
The reduced density matrix is 
\be
\rho_A = \sum_{\mu_i, a_i,c_i} \frac{\mu_1\mu_2}{|G|^2 \mu_3} 
|\omega^{\mu_1\mu_2\bar{\mu}_3}_{a_1a_2a_3}|^2
|\omega^{\mu_1\mu_2\bar{\mu}_3}_{c_1c_2 c_3}|^2
 | \mu_3 ,a_3, c_3\rangle \langle \mu_3, a_3, c_3 | + \cdots
\ee
where $\cdots$ denote components not diagonal in $\mu_3$. Therefore, the probability density of measuring $\mu_3$ is given by
\be \label{prob1}
P(\mu_3) = \sum_{\substack{\mu_1,\mu_2,\\ \mu_1\otimes\mu_2\ni\mu_3}} \frac{\mu_1\mu_2\mu_3}{|G|^2}= 
\sum_{\mu_1,\mu_2} \frac{\mu_1\mu_2\mu_3 N_{\mu_1\mu_2}^{\mu_3}}{|G|^2},
\ee
where in the first equality above we have made use of the fact that 
\be
\sum_{a_i} |\omega^{\mu_1\mu_2\bar{\mu}_3}_{a_1a_2a_3}|^2 = \mu_3.
\ee
Also, the fusion product of of $\mu_1\otimes \mu_2$ is supposed to contain $\mu_3$, but by inserting $N_{\mu_1\mu_2}^{\mu_3}$ corresponding to the fusion matrix, the sum is effectively unconstrained, and hence the second equality. But keep in mind, since we assumed the coupling of representations in this case is multiplicity free, here in fact $N^{\mu_3}_{\mu_1\mu_2}=\delta_{\mu_3\mu_1\mu_2}$, which is either $0$ or $1$. 

Now using the fact that for finite groups,
\be
\begin{aligned}
N_{\mu_1\mu_2}^{\mu_3} &= \frac{1}{|G|} \sum_g \chi^{\mu_1}(g) \chi^{\mu_2}(g) \bar{\chi}^{\mu_3}(g),\\
\sum_\mu \mu\chi^\mu(g)&=|G|\delta_{g,1},
\end{aligned}
\ee 
where $\chi^\mu(g)$ are characters of the representation $\mu$.
Substituting into (\ref{prob1}), we end up with
\be
P(\mu_3) = \frac{\mu_3^2}{|G|},
\ee
which is an answer that makes a lot of sense. To start with, note that $\sum_\mu P(\mu)= 1$. This answer suggests that despite not being to measure in detail the internal states of each given representation $\mu$, the probabilities are in fact sensitive to the dimension of the representations.

Now having done the warmup exercise of a single vertex, we will come to the meat of our discussion. We will consider some arbitrary region in a system with many vertices and links, and try to define an entanglement entropy with the regions beyond. 

Consider region $A$ surrounded by a set of links forming the boundary. The links touching the boundary belonging to $\bar{A}$ would depend on the state that the boundary links of $A$ take due to the Gauss constraint, and we would again like to define an entanglement entropy along the same vein as we have done for the single vertex case. Here, instead of integrating out everything in $\bar{A}$, we could equally ask for the probability of finding the set of protruding links in $\bar{A}$ touching the boundary of $A$ in a particular configuration $\{\mu_l\}$, and that probability would be used for defining the entanglement entropy. The computation pretty much parallels the single vertex case above, with the only new ingredient being that each leg lying along the boundary of $A$ has to connect to two different vertices, and that they form a closed loop, such that one can expect some extra global relation between different touching links in $\bar{A}$, again in complete analogy with the Abelian case. 

Let us assume that there are $L$ boundary links forming a loop, with orientation such that altogether it is an counter-clockwise loop. The $L$ boundary links are separated by $L$ touching links in $\bar{A}$. We note that the external links and the boundary links are meeting at trivalent vertices for simplicity and concreteness. 
\begin{figure}[h]
\centering
\includegraphics[scale=0.8]{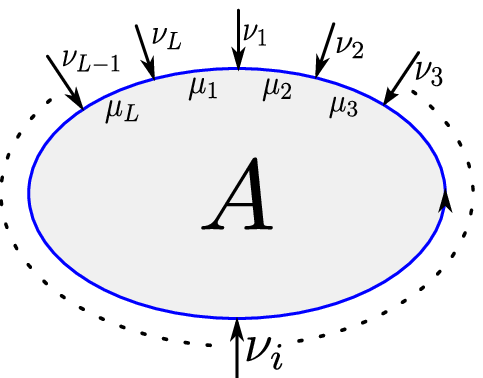}
\caption{}
\end{figure}
Let us allow the links to point toward the loop. Now we would like to compute the probability of finding the set of $L$ external links in any particular configuration denoted by $\{\nu_j|j=1,\cdots L\}$. The representations residing on the boundary of $A$ would be denoted as $\{\mu_i|i=1,\cdots L\}$.

Let us first write down the state
\bea
|\Psi\rangle &&= \frac{1}{|G|^{L/2}}\sum_{\{g_i\}} \prod_i^L \sqrt{\frac{\mu_i}{|G|}} \rho^{\mu_i}(g_i^{-1})_{a_i b_i}|\mu_i,b_i,c_i\rangle \rho^{\mu_i}(g_{i-1})_{c_ia_i} \otimes \prod_j^L \sqrt{\frac{\nu_j}{|G|}}\rho^{\nu_j}(g_j)_{m_jn_j}
|\nu_j,m_j,n_j\rangle\nonumber \\
&&=  \frac{1}{|G|^{L/2}}\sum_{\{g_i\}} \prod_i^L \sqrt{\frac{\mu_i}{|G|}} \rho^{\mu_i}(g_{i-1} g_i^{-1})_{c_i b_i}|\mu_i,b_i,c_i\rangle\otimes \prod_j^L \sqrt{\frac{\nu_j}{|G|}}\rho^{\nu_j}(g_j)_{m_jn_j}
|\nu_j,m_j,n_j\rangle,
\eea
where the indices $i,j$ are taken to be cyclic modulo $L$, i.e., $i=L =0 \pmod L$.

By summing over everything in $A$, we can construct the ``reduced'' density matrix as follows, 
\bea
\rho(\{\nu_i\})=\frac{1}{|G|^L} \sum_{\{g_i\},\{h_i\}, \{m_j,n_j\}} &&\prod_i \frac{\mu_i}{|G|} \chi^{\mu_i}(g_{i-1}g_i^{-1} h_{i}h_{i-1}^{-1}) \otimes \nonumber \\
&& \prod_j \frac{\nu_j}{|G|} \rho^{\nu_j}(g_j )_{m_jn_j} \rho^{\nu_j\,\,*}(h_j)_{m_j n_j}|\nu_j, m_j, n_j\rangle \langle \nu_j, m_j,n_j| + \cdots 
\eea
where we only write down the diagonal components explicitly, and all off-diagonal components are implicitly included in $\cdots$.

Now we would like to compute the probability of finding the external legs in a particular configuration $\{\nu_i\}$.
The probability would involve the sums
\be
\sum_{\{\mu_i\}}\prod_i \frac{\mu_i}{|G|} \chi^{\mu_i}(g_{i-1}g_i^{-1} h_{i}h_{i-1}^{-1})= \prod_i 
\delta_{g_{i-1}g_i^{-1} h_{i}h_{i-1}^{-1},e}
\ee

This product of delta functions imply that
\be
g_{i-1}g_i^{-1}  = h_{i-1}h_{i}^{-1} \Leftrightarrow \, g_i k= h_i,\ \forall i,   
\ee
for any $k\in G$.
Hence, summing over $\{g_i\}$ and $\{h_i\}$ independently can be traded for summing over $\{g_i\}$ and $k$. We are then left with
\bea
P(\{\nu_i\}) =&& \frac{\nu_1\cdots \nu_L}{|G|^{2L}} \sum_{k, \{g_i\}} \prod_i \chi^{\nu_i}(k) \nonumber \\
=&&  \frac{\nu_1\cdots \nu_L}{|G|^{L}} \sum_{k,\{x_i\}} N_{\nu_1\nu_2}^{x_1}N_{x_1\nu_3}^{x_2}\cdots
N_{x_{L-2}\nu_L}^{x_{L-1}} \chi^{x_{L-1}}(k) \nonumber \\
=&& \frac{\nu_1\cdots \nu_L}{|G|^{L-1}}\sum_{\{x_1,...x_{L-2}\}} N_{\nu_1\nu_2}^{x_1}N_{x_1\nu_3}^{x_2}\cdots
N_{x_{L-2}\nu_L}^{x_{L-1}\equiv 1} 
\eea

The $\chi^{\nu_i}$ arises from $\rho^{\nu_i}(g_i)_{m_in_i}\rho^{\nu_i\,*}(g_ik)_{m_in_i}$.
These probabilities clearly sum to 1, again using 
\be
\sum_{\nu} \frac{|\nu|}{|G|}\chi^{\nu}(g)= \delta_{g,e}.
\ee

In the above, we are considering a region $A$ that essentially does not contain an interior, and all its links lie at the boundary. To consider a more general region $A$, there would generally be more links at the boundary compared to the $\nu$ links from $\bar{A}$ touching the boundary of $A$. Nonetheless, as one can see from the above, as we sum over the $\mu_i's$, eventually, all that imposes is 
\be
g_{i-1}^{-1}g_i = h_{i-1}^{-1}h_i,
\ee
despite that now the index $i$ would run from $1$ to $\Lambda$, where $\Lambda \ge L$ in general. 
Therefore, when it comes to the probability distribution of finding the external $\nu$ links
in a particular configuration $P(\{\nu_i\})$, the effect is entirely identical to the above where the interior
is trivial and $\Lambda = L$. 

Let us also remark that if we compare these results with Ref.\onlinecite{Donnelly2014}, there is a reduction in entropy, precisely due to the fact that the internal states are not gauge invariant and are not detectable even experimentally. 

\section{Conclusion}
In this note, we attempt to clarify what the purported "electric entanglement entropy" is measuring, and why it is still an interesting quantity. In the Kitaev model where the physical Hilbert space contains non-Gauss projected states, one recovers exact electric magnetic duality i.e. there is no difference between the so called magnetic center and electric center. We generalize our considerations to non-Abelian gauge theories. 

During the preparation of the manuscript, we become aware of the work \cite{Ghosh2015}, which partially overlaps with our current discussion. In particular, the discussion in \cite{Ghosh2015} emphasizes the fact that the gauge invariant physical spectrum is conveniently embedded in a larger Hilbert space, which is given by the value taken by each link without worrying about gauge fixing/Gauss's constraint. That is precisely the Hilbert space of the Kitaev model that we explored at the beginning of the note.



\acknowledgments
We appreciate William Donnelly, Laurent Freidel, Xi Luo and Rob Myers  for helpful and inspiring discussions. Part of this work was done while LYH was supported by the Croucher Foundation. YW  is supported by the John Templeton Foundation. Research at Perimeter Institute is supported by the Government of Canada through Industry Canada and by the Province of Ontario through the Ministry of Economic Development \& Innovation.

\bibliographystyle{apsrev}

\end{document}